# Multiple magnetic topological phases in bulk van der Waals crystal MnSb$_4$Te$_7$

Shuchun Huan,[1,†] Shihao Zhang,[1,†] Zhicheng Jiang,[2,†] Hao Su[1], Hongyuan Wang,[1] Xin Zhang,[1] Yichen Yang,[2] Zhengtai Liu,[2] Xia Wang,[1,4] Na Yu,[1,4] Zhiqiang Zou[1,4], Dawei Shen[2,5*], Jianpeng Liu[1,3*], Yanfeng Guo[1*]

[1] School of Physical Science and Technology, ShanghaiTech University, Shanghai 201210, China

[2] State Key Laboratory of Functional Materials for Informatics, Shanghai Institute of Microsystem and Information Technology (SIMIT), Chinese Academy of Sciences, Shanghai 200050, China

[3] ShanghaiTech Laboratory for Topological Physics, Shanghai 201210, China

[4] Analytical Instrumentation Center, School of Physical Science and Technology, ShanghaiTech University, Shanghai 201210, China

[5] Center of Materials Science and Optoelectronics Engineering, University of Chinese Academy of Sciences, Beijing 100049, China

The magnetic van der Waals crystals MnBi$_2$Te$_4$/(Bi$_2$Te$_3$)$_n$ have drawn significant attention due to their rich topological properties and the tunability by external magnetic field. Although the MnBi$_2$Te$_4$/(Bi$_2$Te$_3$)$_n$ family have been intensively studied in the past few years, their close relatives, the MnSb$_2$Te$_4$/(Sb$_2$Te$_3$)$_n$ family, remain much less explored. In this work, combining magnetotransport measurements, angle-resolved photoemission spectroscopy, and first principles calculations, we find that MnSb$_4$Te$_7$, the n = 1 member of the MnSb$_2$Te$_4$/(Sb$_2$Te$_3$)$_n$ family, is a magnetic topological system with versatile topological phases which can be manipulated by both carrier doping and magnetic field. Our calculations unveil that its A-type antiferromagnetic (AFM) ground state stays in a $\mathbb{Z}_2$ AFM topological insulator phase, which can be converted to an inversion-symmetry-protected axion insulator phase when in the ferromagnetic (FM) state. Moreover, when this system in the FM phase is slightly carrier doped on either the electron or hole side, it becomes a Weyl semimetal with multiple Weyl nodes in the highest valence bands and lowest conduction bands, which are manifested by the measured notable anomalous Hall effect. Our work thus introduces a new magnetic topological material with different topological phases which are highly tunable by carrier doping or magnetic field.

[†]The authors contributed equally to this work.

[*]Corresponding authors:
dwshen@mail.sim.ac.cn,
liujp@shanghaitech.edu.cn,
guoyf@shanghaitech.edu.cn.

**INTRODUCTION**

Magnetic topological materials have roused a surge of interest in recent years due to their unconventional bulk transport properties [1-9], anomalous surface or edge states [10-22], and the coupling between the magnetic and electronic degrees of freedom, which allow for magnetic control of different topological phases [23-29]. The $MnBi_2Te_4/(Bi_2Te_3)_n$ family is a fertile ground to realize various magnetic topological phases such as axion insulators [25-27], Weyl semimetals (WSMs) [28-30], and quantum anomalous Hall (QAH) insulators [31, 32]. The crystal structure of the parent $MnBi_2Te_4$ is constructed by stacking the "Te-Bi-Te-Mn-Te-Bi-Te" septuple layers (SLs) along the $c$ axis in a triangle lattice via van der Waals (vdW) interaction. $MnBi_2Te_4$ is a bulk topological insulator (TI) with a peculiar A-type antiferromagnetic (AFM) magnetic structure in which the in-plane exchange is ferromagnetic (FM) while the interlayer coupling is AFM [2,5,19,20,33]. This magnetic structure provides access to novel topological phases, such as the QAH effect observed in thin $MnBi_2Te_4$ flake with odd number of SLs under zero magnetic field [31] and an axion insulator state characterized by zero Hall plateau observed in six SLs $MnBi_2Te_4$ [25]. Moreover, the bulk $MnBi_2Te_4/(Bi_2Te_3)_n$ systems exhibit intriguing properties such as the unconventional topological surface states [15, 16, 18-20, 34-37], the complex magnetic phase diagram [27, 38], and the unusual transition behavior with Sb substitutions [39-42].

Although the $MnBi_2Te_4/(Bi_2Te_3)_n$ family have been extensively studied, many issues, such as the zero-field QAH effect and axion insulator state, are still waiting for verifications with more convincing evidences. In this regard, their close relatives, the $MnSb_2Te_4/(Sb_2Te_3)_n$ family, are definitely important and hopeful for the study of the unusual properties of magnetic topological systems. Previous first principles calculations indicate that pristine $MnSb_2Te_4$ is a topologically trivial insulator in both AFM and FM phases, and can be driven into a WSM under substantial compressive strain along the $c$ axis [43], or into a FM TI by the application of external magnetic field [44]. In this work, for the first time we report the successful synthesis of high-quality single crystal MnSb$_4$Te$_7$, a member of the MnSb$_2$Te$_4$/(Sb$_2$Te$_3$)$_n$ family with n = 1, and demonstrate that it exhibits versatile magnetic topological phases based on magnetotransport measurements, first principles calculations, and angle-resolved photoemission spectroscopy (ARPES). Each MnSb$_4$Te$_7$ primitive cell consists of a MnSb$_2$Te$_4$ SL intercalated with an additional Sb$_2$Te$_3$ quintuple layer (QL) as shown in Figure 1(a). The local magnetic moments contributed by the $d$ orbitals of Mn$^{2+}$ ions form long range FM ordering within the SL plane, and the magnetic moments from neighboring SLs are antiferromagnetically ordered along the $c$ axis, establishing the similar A-type AFM state as that of MnBi$_2$Te$_4$, shown schematically in Figure 1(a). The AFM interlayer exchange is weaker compared with that in MnSb$_2$Te$_4$ due to the Sb$_2$Te$_3$ QL intercalation, which results in a lower Néel temperature $T_N$ = 13.5 K. First principles calculations indicate that MnSb$_4$Te$_7$ in the AFM phase is an axion insulator protected by the combination of time-reversal ($\mathcal{T}$) and half lattice translation symmetries ($\tau_{1/2}$), which is associated with gapped topological surface states. Such a phase is also known as a $\mathbb{Z}_2$ AFM topological insulator. The interlayer AFM ordering can be converted into FM ordering under very small vertical magnetic field $\mu_0 H$ > 0.15 T, and then MnSb$_4$Te$_7$ becomes a FM axion insulator protected by inversion symmetry. Moreover, the system in the FM phase would become a WSM upon slight electron or hole doping, which is characterized by notable anomalous Hall effect (AHE) caused by multiple Weyl nodes in the lowest conduction bands and highest valence bands. Our work indicates that MnSb$_4$Te$_7$ is a promising magnetic topological material exhibiting both axion-insulator and Weyl-semimetal phases, which can be tuned by both magnetic field and carrier doping.

The details for crystal growth, magnetization, magnetotransport, ARPES measurements, and first principles calculations of the bulk MnSb$_4$Te$_7$ are presented in Supplementary Information (SI).

## RESULTS AND DISCUSSION

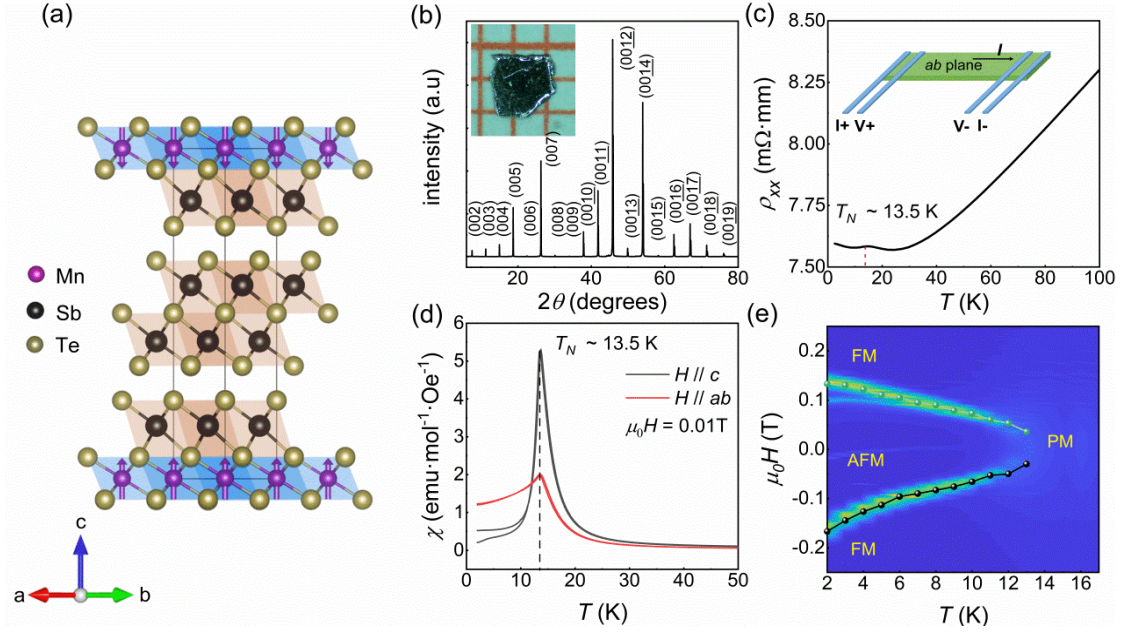

**Fig. 1.** (a) The schematic crystal structure of MnSb$_4$Te$_7$. The blue arrows represent the Mn$^{2+}$ spins in the A-type AFM structure. Green block: edge-sharing SbTe$_6$ octahedra. Blue block: edge-sharing MnTe$_6$ octahedra. (b) The room temperature powder X-ray diffraction peaks from the *ab* plane of MnSb$_4$Te$_7$ crystal. Insert: Image of a typical MnSb$_4$Te$_7$ single crystal synthesized in this work. (c) The temperature dependent of transverse resistivity $\rho_{xx}$ at $\mu_0 H = 0$ T measured from 2 K to 100 K. The sketch for the four-probe measurements configuration is inserted. (d) The temperature dependence of susceptibility $\chi$ under $\mu_0 H = 0.01$ T for $H // c$ and $H // ab$ plane. (e) The magnetic phase diagram of MnSb$_4$Te$_7$. The solid dots and line with solid dots represent the critical magnetic fields as a function of temperature.

Figure 1(b) shows the room temperature Bragg reflections from the *ab* plane of a typical MnSb$_4$Te$_7$ single crystal, which can be nicely indexed by using the MnBi$_4$Te$_7$ structure model. A picture of a typical MnSb$_4$Te$_7$ single crystal is shown as an insert of Figure 1(b), with the dimensions of about $2 \times 2 \times 0.2$ mm$^3$ and the flat surface corresponding to the *ab* plane.

The resistivity and magnetic properties of MnSb$_4$Te$_7$ are depicted in Figures 1(c)-(e), respectively. Seen from the temperature dependence of longitudinal resistivity $\rho_{xx}$ at $\mu_0H = 0$ T with $I$ // $ab$ plane presented in Figure 1(c), it decreases linearly with decreasing the temperature to ~ 25 K, and then slightly increases below 20 K upon further cooling. The slight upturn of $\rho_{xx}$ at low temperature likely originates from the enhanced scattering amplitudes due to the critical magnetic fluctuations approaching the Néel temperature, which is commonly observed in low dimensional magnetic systems. An abrupt drop of $\rho_{xx}$ is observed at 13.5 K, consistent with the AFM ordering temperature manifested by the magnetic susceptibility presented in Figure 1(d). Seen in both zero-field-cooled (ZFC) and field-cooled (FC) curves measured at $\mu_0H = 0.01$ T with $H$ // $c$ axis and $H$ // $ab$ plane, respectively, abrupt transitions around $T_N = 13.5$ K are visible, indicating an intrinsic long range AFM ordering, which is similar to that observed in MnBi$_4$Te$_7$ [6,9,25]. The magnetic susceptibility decreases dramatically bellow $T_N$ for $H$ // $c$, but only slightly decreases for $H$ // $ab$, unveiling an anisotropic AFM exchange with the $c$ axis being the magnetic easy axis. When $H$ is set along the $c$ axis, the ZFC and FC curves are slightly separated from each other, implying possible hysteresis behavior [See details in Figure S2]. The weak interlayer AFM coupling can be easily converted into FM order by vertical magnetic field. The critical magnetic fields for such AFM-FM transition as a function of temperature are marked by the blue (for positive $H$) and black (for negative $H$) dots in Figure 1(e), which decreases with increasing temperature. Eventually the system enters the paramagnetic phase for $T > T_N$, as marked by "PM" in Figure 1(e).

The MnSb$_4$Te$_7$ system obeys the symmetries of $P$-3$m$1 space group, which include inversion symmetry (inversion center is located at the Mn atom) as shown Figure 1(a). First principles calculations are employed to investigate the electronic structure of MnSb$_4$Te$_7$ with four different magnetic structures. The detailed methods can be found in the SI. First, we investigate the AFM ground state. Figure 2(a) reveals that the MnSb$_4$Te$_7$ in the AFM structure has a direct gap of 75 meV. When the system is in the FM structures, the bulk systems have indirect gaps around 60 meV as shown in Figures 2(b)-2(d), for the FM magnetizations pointing along *z*, *x*, and *y* directions, respectively. The three FM phases with magnetizations pointing along *x*, *y*, and *z* directions are denoted by "FMx", "FMy" and "FMz" in the following discussions. The energy bands near the Fermi level are mainly contributed by the Sb *p* and Te *p* orbitals with substantial spin-orbit coupling (SOC). The localized 3*d* orbitals from the Mn$^{2+}$ ions contribute to the local magnetic moments, which are coupled to the conduction electrons through effective Zeeman couplings, thus combining SOC with magnetism in the electronic degrees of freedom and enabling the realization of different magnetic topological phases in the system.

**Table I.** The symmetries of bulk MnSb$_4$Te$_7$ in different magnetic states.

|     | $P$ | $\mathcal{T}\tau_{1/2}$ | $M_x$ | $C_{3z}$ | $C_{2x}$ |
|-----|-----|------------------------|-------|----------|----------|
| AFM | ✓ | ✓ | ✗ | ✓ | ✗ |
| FMx | ✓ | ✗ | ✓ | ✗ | ✓ |
| FMy | ✓ | ✗ | ✗ | ✗ | ✗ |
| FMz | ✓ | ✗ | ✗ | ✓ | ✗ |

The bulk MnSb$_4$Te$_7$ crystal in the non-magnetic phase has $P$, $M_x$, $C_{3z}$, and $C_{2x}$ symmetries. The AFM state doubles the primitive cell along the *c* axis, which breaks the $M_x$ and $C_{2x}$, but preserves $P$ symmetry and an additional non-symmorphic symmetry which combines $\mathcal{T}$ and $\tau_{1/2}$. Inversion symmetry is also preserved in all the FMx, FMy, and FMz phases. The symmetries of bulk MnSb$_4$Te$_7$ with the four different magnetic configurations are enumerated in Table I.

Because the MnSb$_4$Te$_7$ system preserves inversion symmetry in all the four magnetic configurations, we can use the symmetry indicator $\mathbb{Z}_4$ invariant to characterize the topological characters of the different magnetic topological phases [45-48]. Here the $\mathbb{Z}_4$ invariant can be defined as $\mathbb{Z}_4 = \sum_{k=1}^{8}(n_k^+ - n_k^-)/2 \ mod \ 4$, where $n_k^+/n_k^-$ is the number of occupied states with even/odd parity at the inversion-invariant momenta $\mathbf{k}$. $\mathbb{Z}_4 = 1$ or 3 indicates a WSM phase, while $\mathbb{Z}_4 = 2$ means that the system is an axion insulator given that the Chern numbers on all the 2D $\mathbf{k}$ planes in the Brillouin zone are zeros. The parity analysis reveals that all these magnetic states are axion insulators with quantized bulk orbital magnetoelectric coupling and gapped topological surface states [49-54], and the parities of the occupied bands at the inversion-invariant $\mathbf{k}$ points are shown in Table. S2 of the SI. The hybrid Wannier charge center evolution (also known as the "Wilson loops") [55, 56] in the four magnetic states in the $k_z = 0$ plane also indicates that the system in different magnetic configurations exhibits nontrivial topological properties as characterized by the winding pattern of the Wilson loops, which are shown in Figure S5 of SI. It is worthwhile to note that in addition to inversion symmetry, MnSb$_4$Te$_7$ in the AFM phase also has $\mathcal{T}\tau_{1/2}$ symmetry, which allows for a $\mathbb{Z}_2$ classification. Thus, MnSb$_4$Te$_7$ in the AFM phase can also be classified as a $\mathbb{Z}_2$ AFM TI [57]. Different from the MnBi$_2$Te$_4$ system which only stays in the axion-insulator state in the AFM configuration but becomes a WSM in the FM configuration [23,27], MnSb$_4$Te$_7$ system retains the robust axion-insulator state regardless of magnetic ordering or magnetic orientation. More interestingly, in the FM configurations the MnSb$_4$Te$_7$ system behaves as a WSM upon slight electron or hole doping, because there are multiple Weyl nodes in the lowest conduction bands and highest valence bands, and the positions of these Weyl nodes can be moved by changing the magnetization direction. Such WSM phase introduced by carrier doping will be discussed in detail in the following. We see that the MnSb$_4$Te$_7$ system in the FM configurations realizes a unique phase with co-existing axion-insulator and Weyl-semimetal states, the transition between which can be electrically tuned by slight carrier doping.

The MnSb$_4$Te$_7$ system has two types of surface terminations on the (001) plane: MnSb$_2$Te$_4$ SLs and Sb$_2$Te$_3$ QLs. Our calculation results shown in Figure S6 of SI indicate that there is a Dirac-like crossing near the Fermi level (E$_F$) in the spectrum of surface states with SL termination, but it becomes parabolic energy bands with larger band gap when the termination turns to the QL termination. This is similar to what has been observed in MnBi$_4$Te$_7$ [12,14,15,34]: the first type of surface states with SL termination are genuine Dirac surface states which are gapped out due to the Zeeman coupling to the surface magnetizations; while the second type of surface states with QL termination results from the hybridization between the states in the QLs and SLs. More details about the calculated surface states of MnSb$_4$Te$_7$ can be found in SI.

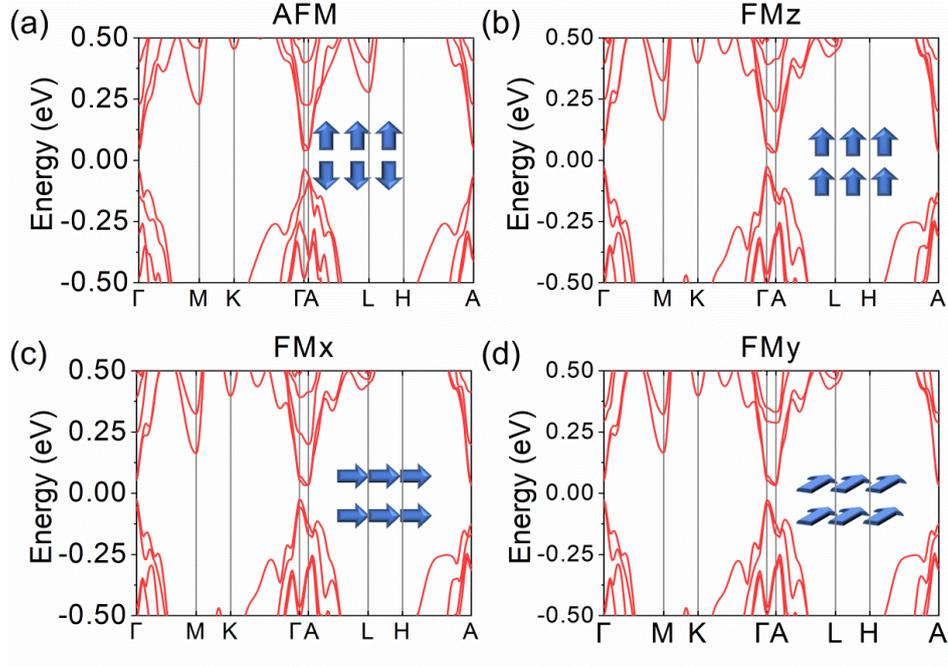

**Fig. 2.** The band structures of (a) A-type AFM state, (b) FM state with spin aligned along *z* direction (FMz), (c) FM state with *x*-direction spin (FMx), and (d) FM state with *y*-direction spin (FMy) of bulk MnSb$_4$Te$_7$ with considering the spin-orbital coupling effect. Here Fermi levels are set to zero and spin directions are remarked by blue arrows.

We continue to discuss the magnetotransport properties of MnSb$_4$Te$_7$ in different magnetic configurations. The synthesized bulk MnSb$_4$Te$_7$ crystals are slightly hole doped which behave as metals in transport. Figure 3(a) shows the magnetic field $H$ dependence of $\rho_{xx}(H)$. The AFM states are manifested by the resistivity plateaus for weak magnetic fields $H < H_c$ ($\mu_0 H_c \sim$ 0.1 - 0.2 T) at temperatures lower than $T_N$. Spin-flop transitions are observed for $H > H_c$ and $T < T_N$, as characterized by the abrupt drop of $\rho_{xx}$ when $H > H_c$, which is in good accordance with the previous magnetization measurements that suggest an AFM-FM transition at $H_c$. When the magnetic field is further enhanced, $\rho_{xx}$ is slightly decreased implying that the local moments of the Mn$^{2+}$ ions are fully saturated. The field dependence of the anomalous Hall resistivity $\rho_{xy}^A$ is displayed in Figure 3(b). Clearly, when $T < T_N$ and $H > H_c$, the MnSb$_4$Te$_7$ system exhibits significant AHE with notable hysteresis loops shown in Figure 3(b). The magnitude of $\rho_{xy}^A$, the area of the hysteresis loops, and the magnetization $M(H)$ all gradually diminish with the increase of temperature for $H > H_c$ and $T < T_N$, and they eventually (almost) vanish for $T > T_N$ (see details in Figure S2 of SI). The converted anomalous Hall conductivity (AHC) $\sigma_{xy}^A$ ($= -\rho_{xy}^A/(\rho_{xy}^{A~2} + \rho_{xx}^2)$) is presented in Figure 3(c). To determine the dominant mechanism for the AHC, the $\rho_{xy}^A$ vs. $\rho_{xx}$ relationship is fitted using the method described in Ref. [1]. As shown in Figure 3(d), $\rho_{xy}^A/(M\rho_{xx})$ is linearly dependent on $\rho_{xx}$ when $\rho_{xx} < 7.6$ mΩ·mm, suggesting that the dominant contribution to AHC in MnSb$_4$Te$_7$ is from the intrinsic Berry curvatures of the band structures. At 2 K, $\rho_{xy}^A$ is extracted to be 54 μΩ·mm and $\sigma_{xy}^A$ is 0.99 Ω$^{-1}$ mm$^{-1}$, which is rather close to the theoretical value 1.12 Ω$^{-1}$ mm$^{-1}$ as discussed in detail later. It should be noted that the low temperature upturn of $\rho_{xx}$ is possibly due to slightly enhanced electron-electron interactions [58, 59], which does not change the intrinsic nature of the AHE.

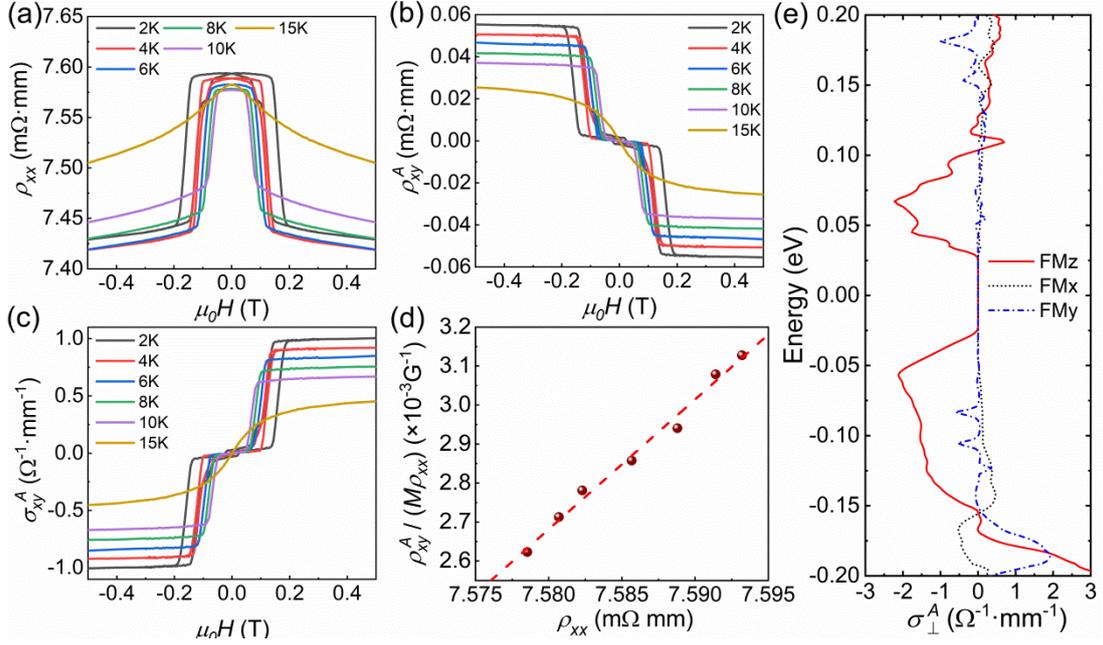

**Fig. 3.** (a) Magnetoresistance $\rho_{xx}$ with $I // ab$ plane and $H // c$ at various temperatures. (b) The anomalous hall resistivity $\rho_{xy}^A$. (c) The anomalous hall conductivity $\sigma_{xy}^A$ at various temperatures. (d) The plot of $\rho_{xy}^A/(M\rho_{xx})$ vs. $\rho_{xx}$. (e) The calculated anomalous Hall conductivities of MnSb$_4$Te$_7$ for the three different ferromagnetic states as a function of varied Fermi energy, where the actual Femi energy is set as zero.

We also have calculated the AHC of the three FM phases of MnSb$_4$Te$_7$ with 151×151×151 $k$-points mesh as shown in Figure 3(e). Near the $E_F$, the AHC of FMz state is larger than those of the other two FM states. It should be noted that the calculated AHC component is always in the plane perpendicular to the magnetization direction. When the system is slightly hole doped, there are four Weyl nodes of FMx state in the -0.15 ~ 0 eV range located at ±(±0.003, 0.049, 0.066) ang$^{-1}$ (0 is set to be the middle of the energy gap), while there are only two Weyl nodes in the FMz state in the same energy range located at ±(-0.003, 0.049, 0.066) ang$^{-1}$ due to the lack of $M_x$ mirror symmetry in FMz phase. The four Weyl nodes of FMx state are almost in the $k_x = 0$ plane, thus contribute little to the $yz$ component of the AHC. But the Weyl nodes of FMz state are well separated from the $k_z = 0$ plane, so we can see the remarkable AHC $\sigma_{xy}^A$ compared to other magnetic states.

To confirm the non-trivial topological properties of MnSb$_4$Te$_7$, we used the synchrotron-based ARPES to probe its low-energy band structure, as shown in Figure

4. By changing the photon energy in an extensive range of 50 ~ 90 eV, which covers more than one whole Brillouin zone, we can distinguish unaltered Dirac surface states from the bulk ones, as marked by red arrows in Figure 4 (a). These ARPES spectra demonstrate the prominent p-type carrier dosage in MnSb$_4$Te$_7$. Through detailed comparison with the calculated band dispersion along the Γ-K direction [Figure 4(b)], we can evaluate that the Fermi level of the as-grown MnSb$_4$Te$_7$ is located ~180 meV below the Dirac point (DP), as illustrated in Figure 4(c). We note that the calculated $\sigma_{xy}^A$ based on this estimation of E$_F$ is 1.12 Ω$^{-1}$·mm$^{-1}$, which is in remarkable agreement with the transport result taken at 2 K (0.99 Ω$^{-1}$·mm$^{-1}$). To further observe the Dirac surface states, we tried to tune its chemical potential by Bi substitution for Sb in MnSb$_4$Te$_7$. As shown in Figure 4(d), with ~15% Bi substitution, the E$_F$ of Mn(Bi$_{0.15}$Sb$_{0.85}$)$_4$Te$_7$ is discovered to shift up by ~120 meV, while leaving the main band dispersion unchanged. Furthermore, we can extrapolate the Dirac band to its apex to obtain the position of DP, which is only 60 ± 10 meV above E$_F$, in line with our previous estimation for as-grown MnSb$_4$Te$_7$. Potassium surface doping has been tried as well, while the chemical potential does not show significant movement, which might be attributed to the large density of states close to E$_F$ (see details in Figure S10). In addition, can further prove the existence of a surface Dirac cone in the vicinity of E$_F$.

The Dirac cone is further shown by constant-energy cuts of Mn(Bi$_{0.15}$Sb$_{0.85}$)$_4$Te$_7$ [Figure 4(e)]. Also, the experimental petal-like Fermi surface is in a good agreement with the calculation for SL termination [Figure S8(a)]. Nevertheless, MnSb$_4$Te$_7$ is expected to expose two kinds of terminations after cleaving, namely SL (MnSb$_2$Te$_4$) and QL (Sb$_2$Te$_3$). We have well scanned the whole sample cleaved surface with our 30 μm × 30 μm synchrotron beam, while we could not distinguish between these two terminations, which is distinct from the case of MnBi$_4$Te$_7$ [14, 15, 34]. This might result from their rather similar band structure below the Dirac point according to our calculations [Figure S8 of SI]. Moreover, similar to MnBi$_4$Te$_7$, we as well find a dispersionless feature with 90 eV photons at the binding energy of ~ 3 eV, corresponding to the contributions of Mn 3$d$-states (Figure S9).

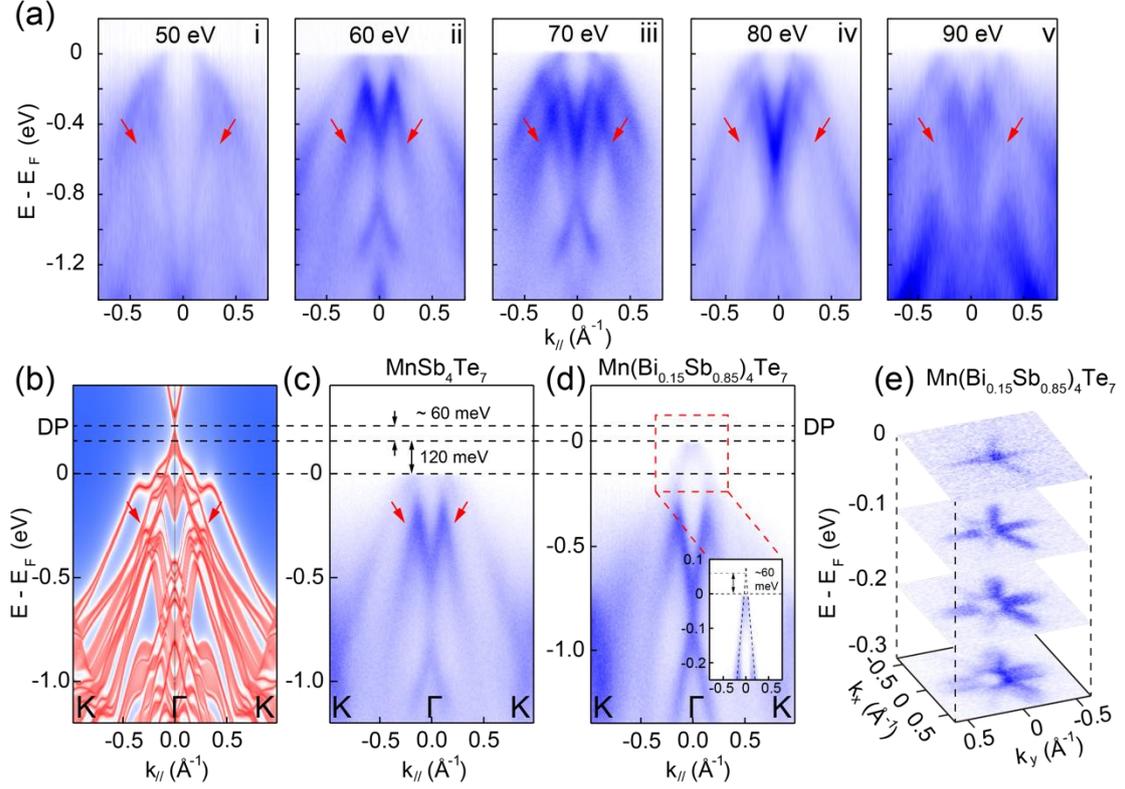

**Fig. 4.** (a) ARPES intensity plot along the Γ-K direction measured at $h\nu$ = 50, 60, 70, 80 and 90 eV at 15 K. (b) Calculated band structure along Γ-K direction of SL termination. ARPES intensity mapped along the Γ-K direction of (c) MnSb$_4$Te$_7$ and (d) Mn(Bi$_{0.15}$Sb$_{0.85}$)$_4$Te$_7$ collected at 60 eV. (e) Constant-energy surface of the Dirac cone cut for Mn(Bi$_{0.15}$Sb$_{0.85}$)$_4$Te$_7$.

## SUMMARY

To summarize, in this work we report the first successful growth of high-quality MnSb$_4$Te$_7$ single crystal, and have studied its magnetic topological properties combining transport measurements, first principles calculations, and ARPES studies.

We deduce that bulk MnSb$_4$Te$_7$ would stay in axion-insulator states for both FM and AFM configurations, which are associated with topologically nontrivial surface states that are confirmed by our ARPES measurements. Moreover, there are multiple Weyl nodes in the highest valence bands and lowest conduction bands in the FM phase such that the system behaves as a Weyl semimetal upon slight carrier doping, which is characterized by notable anomalous Hall effect. Therefore, the MnSb$_4$Te$_7$ in the FM configuration realizes a unique phase with co-existing axion-insulator and Weyl-semimetal states, and the transition between the two topological states can be tuned by carrier doping. Our work is a significant step forward in understanding the intriguing physical properties of magnetic topological materials, and will provide useful guidelines for future experimental and theoretical studies in this field.

**ACKNOWLEDEMENTS**

The authors acknowledge the support by the Major Research Plan of the National Natural Science Foundation of China (No. 92065201), the National Natural Science Foundation of China (Grant Nos. 11874264 and U2032208), and the National Key R&D Program of the MOST of China (Grant No. 2016YFA0300204). S. Zhang and J. Liu acknowledge the start-up grant of ShanghaiTech University and the National Key R & D program of China (grant no. 2020YFA0309601). Y. F. Guo acknowledges the start-up grant of ShanghaiTech University and the Program for Professor of Special Appointment (Shanghai Eastern Scholar). Part of this research used Beamline 03U of the Shanghai Synchrotron Radiation Facility, which is supported by ME2 project under contract No. 11227902 from National Natural Science Foundation of China. The authors also thank the support from the Analytical Instrumentation Center (SPST-AIC10112914), SPST, ShanghaiTech University.

# Supplementary Information

# Multiple magnetic topological phases in bulk van der Waals crystal MnSb$_4$Te$_7$

Shuchun Huan,[1,†] Shihao Zhang,[1,†] Zhicheng Jiang,[2,†] Hao Su,[1] Hongyuan Wang,[1] Xin Zhang,[1] Yichen Yang,[2] Zhengtai Liu,[2] Xia Wang,[1,4] Na Yu,[1,4] Zhiqiang Zou,[1,4], Dawei Shen,[2,5*] Jianpeng Liu,[1,3*] Yanfeng Guo[1*]

[1] School of Physical Science and Technology, ShanghaiTech University, Shanghai 201210, China

[2] State Key Laboratory of Functional Materials for Informatics, Shanghai Institute ofMicrosystem and Information Technology (SIMIT), Chinese Academy ofSciences, Shanghai 200050, China

[3] ShanghaiTech Laboratory for Topological Physics, Shanghai 201210, China

[4] Analytical Instrumentation Center, School of Physical Science and Technology, ShanghaiTech University, Shanghai 201210, China

[5]Center of Materials Science and Optoelectronics Engineering, University of Chinese Academy of Sciences, Beijing 100049, China

[†]These authors contributed equally to this work.

*Corresponding authors:

dwshen@mail.sim.ac.cn,

liujp@shanghaitech.edu.cn,

guoyf@shanghaitech.edu.cn.

## a. Crystal growth and characterizations

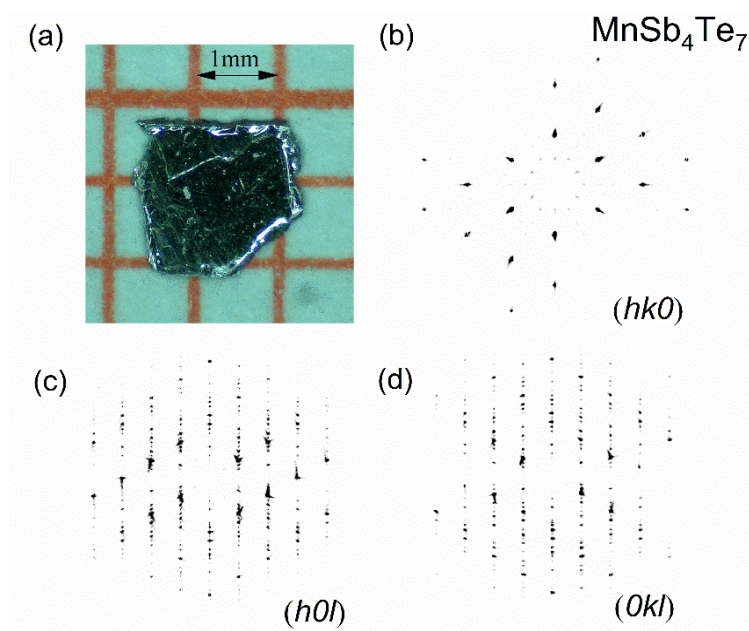

**Fig. S1.** (a) A typical picture for obtained MnSb$_4$Te$_7$ single crystal. (b) – (c) Single crystal X-ray diffraction patterns in the reciprocal space along the (*0kl*), (*h0l*) and (*hk0*) directions for MnSb$_4$Te$_7$ measured at 298 K.

The MnSb$_4$Te$_7$ single crystals were grown by using the self-flux method. Starting materials of Mn (99.95%, aladdin), Sb (99.999%, aladdin) and Te (99.9999%, aladdin) blocks were mixed in a molar ratio of 1: 10: 15 and placed into an alumina crucible which was then sealed into a quartz tube in vacuum. The assembly was heated in a furnace up to 750 $^{o}$C within 10 hrs, kept at the temperature for 15 hrs, and then slowly cooled down to 613 $^{o}$C at a temperature decreasing rate of 2 $^{o}$C/h. The excess melt components were removed at this temperature by quickly placing the assembly into a high-speed centrifuge and black crystals with shining surface in a typical dimension of 2 ×2 ×0.2 mm$^3$, shown by the insert picture as Figure S1(a).

The phase and quality of the single crystals used in this work were examined on a Bruker D8 Venture single crystal X-ray diffractometer (SXRD) with Mo $K_{\alpha 1}$ ($\lambda$ = 0.71073Å) and Bruker D8 Advance powder X-ray diffractometer (PXRD) with Cu $K_{\alpha 1}$ ($\lambda$ = 1.54184Å) at 298 K. The PXRD data are presented in the main text. The SXRD diffraction patterns of MnSb$_4$Te$_7$ shown in Figures S1(b)-1(d) could be satisfyingly indexed on the basis of a trigonal structure with the lattice parameters with $a = b$ = 4.25 Å, $c$ = 23.76 Å, $\alpha = \beta$ = 90 ° and $\gamma$ = 120 ° in the space group $P$-3$m$1 (No. 164). The refinement results are summarized in Table S1. The prefect reciprocal space lattice indicates the high quality of our single crystal sample.

**Table S1. Refinement results for MnSb$_4$Te$_7$ based on the single crystal x-ray diffraction data collected at room temperature.**

| Atom | site | $x$ | $y$ | $z$ | $U$ |
|---|---|---|---|---|---|
| Mn | 1b | 1 | 0 | 0 | 0.0171 |
| Sb1 | 2d | 0.3333 | 0.6667 | 0.4167 | 0.0424 |
| Sb2 | 2d | 0.3333 | 0.6667 | 0.1579 | 0.0469 |
| Te1 | 1b | 0 | 1 | 0.5 | 0.0362 |
| Te2 | 2d | 0.6667 | 0.3333 | 0.3457 | 0.0372 |
| Te3 | 2c | 0 | 1 | 0.2271 | 0.0369 |
| Te4 | 2d | 0.6667 | 0.3333 | 0.0711 | 0.0384 |

### b. Magneto-transport measurements

The magnetizations were measured by using a commercial Magnetic Property Measurement System (MPMS-3). The alternating current magnetization was measured at the magnetic field $\mu_0H$ = 10 Oe and the frequency of 273Hz. Magnetotransport measurements, including the resistivity, magnetoresistance and Hall effect measurements, were carried out in commercial DynaCool Physical Properties

Measurement System from Quantum Design. The resistivity and magnetoresistance were measured in a four-probe configuration and the Hall effect measurement was using a standard six-probe method.

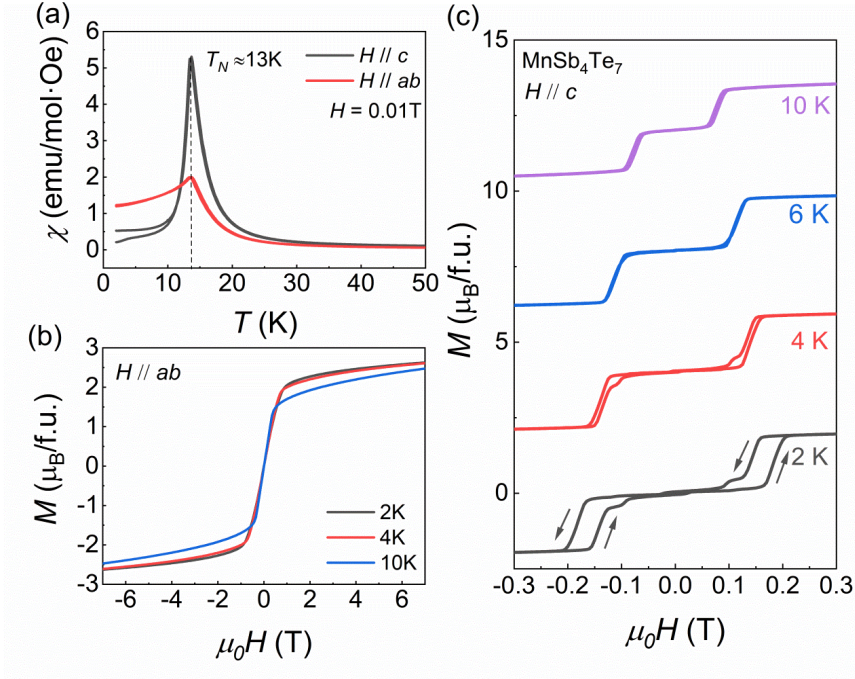

**Fig. S2.** (a) The temperature dependence of magnetic susceptibility $\chi$ under $\mu_0 H = 0.01$ T for $H \mathbin{/\mkern-2mu/} c$ and $H \mathbin{/\mkern-2mu/} ab$ plane. (b) Isothermal magnetizations at 2 K, 4 K and 10 K for $H \mathbin{/\mkern-2mu/} ab$. (c) Isothermal magnetizations at 2 K, 4 K, 6 K and 10 K for $H \mathbin{/\mkern-2mu/} c$.

As mentioned in the main text, the intrinsic long range AFM order could be identified in the magnetic susceptibility curves. when $H \mathbin{/\mkern-2mu/} c$, the ZFC and FC curves separate slightly at low temperature, indicating the possibility of hysteresis loops, which could by further demonstrated by the hysteresis loops of isothermal magnetizations as shown in Figure S2(c), where a first-order spin flip transition with hysteresis at 2 K beginning at $\mu_0 H = 0.16$ T and vanishing at $\mu_0 H = 0.21$ T is visible. At $\mu_0 H = 0.21$ T, the system enters into the forced FM state, which is similar as the case of MnBi$_4$Te$_7$ [1, 2]. Comparing with the MnSb$_2$Te$_4$, the smaller saturation magnetic field indicates the weaker $Mn^{2+}$ - $Mn^{2+}$ interlayer AFM exchange. While increasing the temperature to 10 K, the hysteresis loop gradually diminishes to be nearly zero, but the spin-flip transition is still observable below $T_N$. The saturation moment at 2 K under $\mu_0 H = 0.3$ T is $1.93\mu_B/Mn^{2+}$, which is smaller than the value of $3.5\mu_B/Mn^{2+}$ of $MnBi_4Te_7$ [1]. The reason may lie in that the Mn disorders and mixed location of Mn and Sb atoms [3], or due to the enhanced hole-carrier mediated Ruderman-Kittle-Kasuya-Yosida (RKKY) interaction which could give rise to magnetic frustration [4-6]. Figure S2(b) displays the isothermal magnetizations with *H // ab*, showing that the saturation field is about 1 T with the almost similar saturation moment, indicating that the *c* axis is the magnetic easy axis.

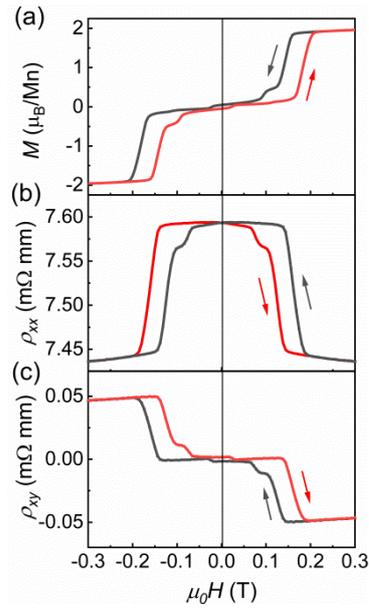

**Fig. S3.** Magnetotransport properties of bulk $MnSb_4Te_7$ single crystal. (a) The field dependence of magnetization *M*, (b) transverse magnetoresistance $\rho_{xx}$, and (c) Hall resistivity $\rho_{xy}$ at 2 K with *H // c* and *I // ab* plane.

The magnetotransport properties are toughly related with the magnetic structure. For $MnSb_4Te_7$, as displayed in Figure S3, $\rho_{xx}(H)$ and $\rho_{xy}(H)$ follow the same hysteresis loop as that in *M(H)* with *H // c* and *I // ab* plane, where the red and black arrows represent the paths for changing the magnetic field. The MnSb$_4$Te$_7$ single crystal used for the magnetotransport measurements has the thickness of ~ 0.1 mm. The magnetic field dependent $\rho_{xx}(H)$ and $\rho_{xy}(H)$ measured in the magnetic field range of $\mu_0H$ = -1.5 T – 1.5 T are shown in Figures S4(a) and (b), respectively.

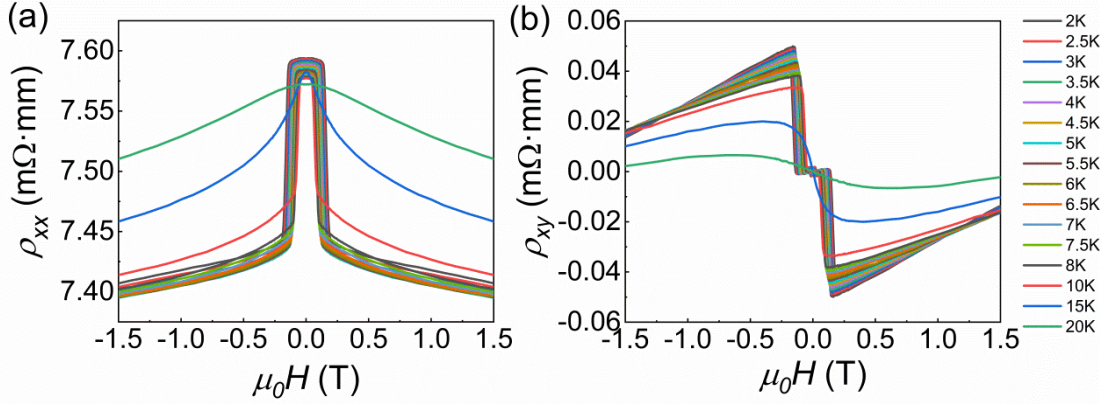

**Fig. S4.** Magnetotransport properties of bulk MnSb$_4$Te$_7$ single crystal. (a) Magnetoresistance $\rho_{xx}$ with $I // ab$ plane and $H // c$ at various temperatures. (b) Hall resistivity $\rho_{xy}$ measured at the same temperatures as in (a).

### c. First-principle calculations

The first-principle calculations were carried out in the framework of the generalized gradient approximation (GGA) functional [7] of the density functional theory through employing the Vienna *ab initio* simulation package (VASP) [8] with projector augmented wave method [9]. The SOC effect was included in calculations. Considering the transitional metal Mn element, U = 3 eV are used in all calculations. The Bloch states are projected to the Wannier functions [10, 11] to build the tight-binding Hamiltonian, and then use irvsp program [12] and WannierTools package [13] to calculate the anomalous Hall conductivity.

We also consider the slab with few layers which keep antiferromagnetic and have two same terminations, and their energy bands are present in the Figure S7. We can see the band gap of about 20 meV will disappear while the thickness of antiferromagnetic slab is changed from five layers to nine layers.

**Table S2. The numbers of occupied states with even/odd parity at the inversion-invariant momenta of bulk MnSb$_4$Te$_7$ at the four different magnetic states.**

| Time-reversal invariant momenta | Number of occupied bands with even parity (+) | Number of occupied bands with even parity (-) |
|---|---|---|
| AFM | | |
| (0,0,0) | 76 | 74 |
| (1/2,0,0), (0,1/2,0), (1/2,1/2,0) | 74 | 76 |
| (0,0,1/2) | 75 | 75 |
| (1/2,0,1/2), (0,1/2,1/2), (1/2,1/2,1/2) | 75 | 75 |
| FMz | | |
| (0,0,0) | 37 | 38 |
| (1/2,0,0), (0,1/2,0), (1/2,1/2,0) | 35 | 40 |
| (0,0,1/2) | 39 | 36 |
| (1/2,0,1/2), (0,1/2,1/2), (1/2,1/2,1/2) | 39 | 36 |
| FMx | | |
| (0,0,0) | 37 | 38 |
| (1/2,0,0), (0,1/2,0), (1/2,1/2,0) | 35 | 40 |
| (0,0,1/2) | 39 | 36 |
| (1/2,0,1/2), (0,1/2,1/2), (1/2,1/2,1/2) | 39 | 36 |
| FMy | | |
| (0,0,0) | 37 | 38 |
| (1/2,0,0), (0,1/2,0), (1/2,1/2,0) | 35 | 40 |
| (0,0,1/2) | 39 | 36 |
| (1/2,0,1/2),(0,1/2,1/2),(1/2,1/2,1/2) | 39 | 36 |

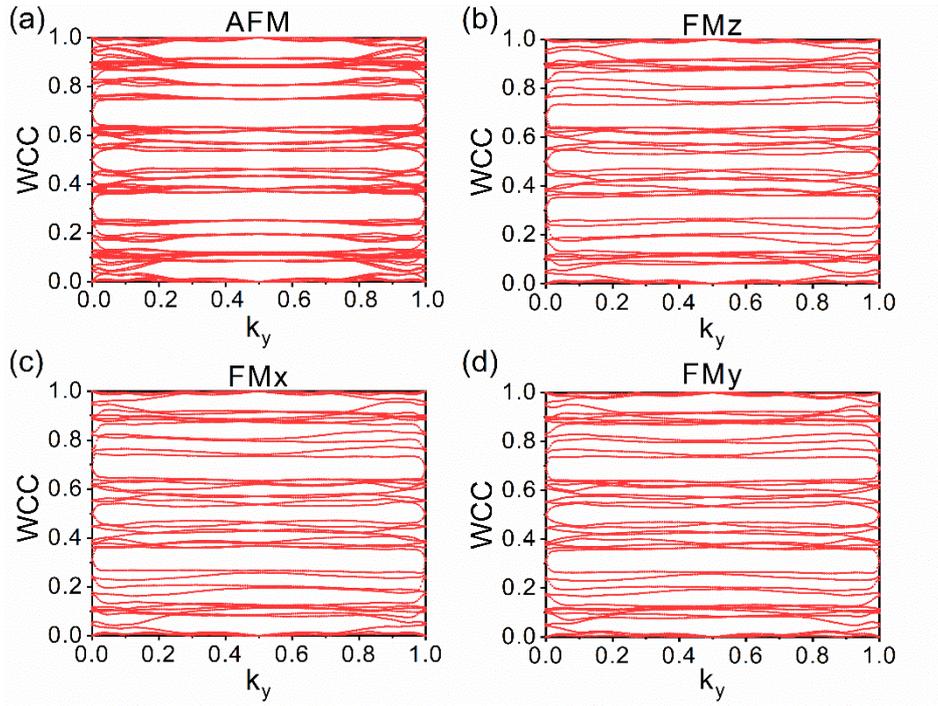

**Fig. S5.** The Wannier charge center (WCC) evolution in four magnetic states in the $k_z = 0$ plane.

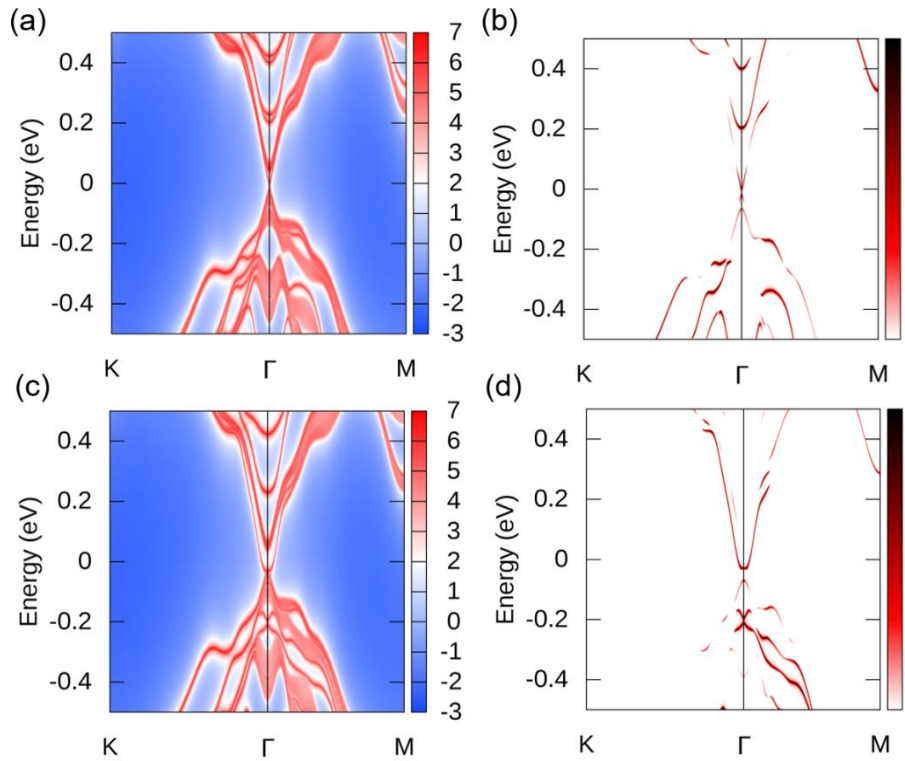

**Fig. S6.** (a) The calculated k-E map of $MnSb_4Te_7$ with the $MnSb_2Te_4$ SL termination. The spectrum of only surface states is present in the (b). (c) The calculated k-E map of $MnSb_4Te_7$ with the $Sb_2Te_3$ QL termination. The spectrum of only surface states is in the (d).

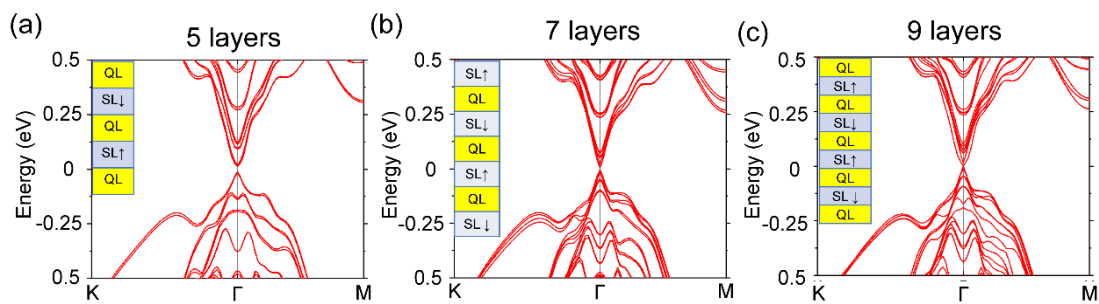

**Fig. S7.** The calculated energy bands of various few layers of $MnSb_4Te_7$ in the AFM configurations.

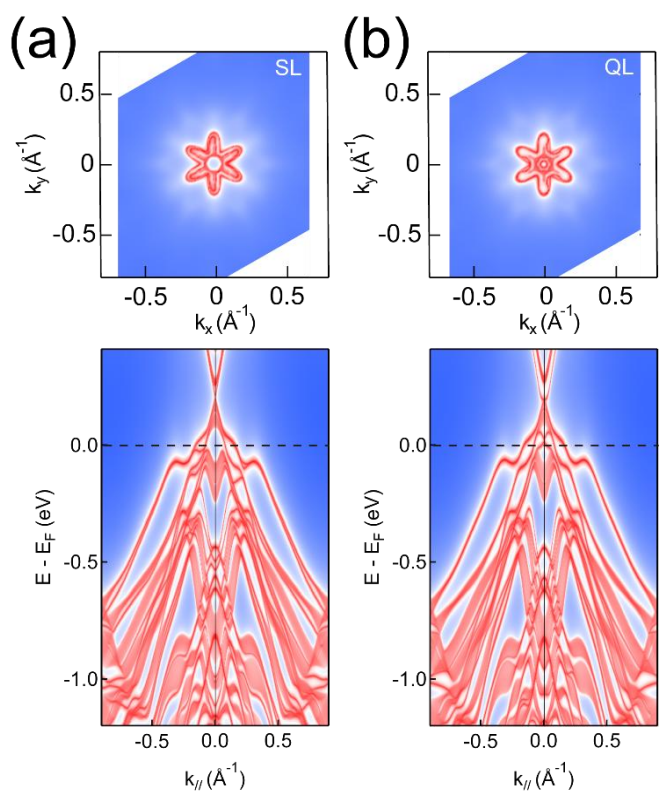

**Fig. S8.** Calculated Fermi surface and band structure of $MnSb_4Te_7$ system with (a) SL and (b) QL terminations. The directions of the energy bands are along -K-Γ-K.

### d. The ARPES measurements

ARPES measurements were performed at beam line 03U of Shanghai Synchrotron Radiation Facility (SSRF). The samples were cleaved in situ and measured under ultrahigh vacuum below $8 \times 10^{-11}$ Torr. Data were collected by Scienta DA30 analyzer. The total energy resolutions are 15 ~ 18 meV with photon energy 50 ~ 90 eV and angle resolution is 0.2°.

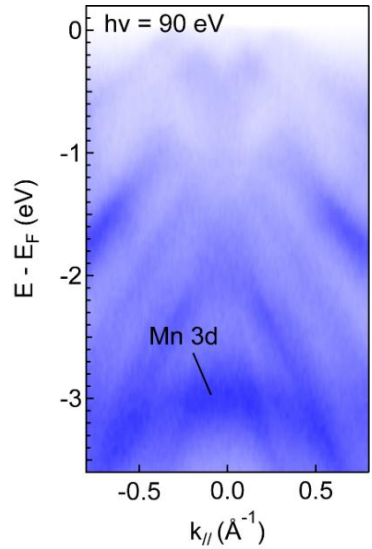

**Fig. S9.** Valence band structure of MnSb$_4$Te$_7$ collected at 15 K with photon energy of 90 eV.

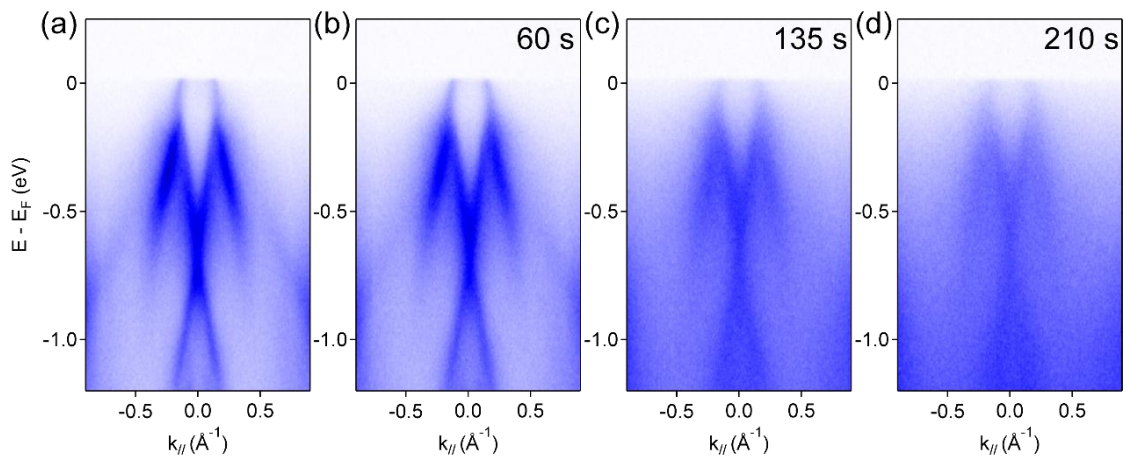

**Fig. S10.** ARPES intensity map of potassium doped MnSb$_4$Te$_7$ with different time: (a) 0s; (b) 60s;

(c) 135s; (d) 210s.